\documentclass{article}
\usepackage[accepted]{mlsys2021}

\usepackage{microtype}
\usepackage{graphicx}
\usepackage{subfigure}
\usepackage{booktabs} 
\usepackage{makecell}
\usepackage{pifont}
\usepackage{caption}
\usepackage{xcolor}
\usepackage{listings}
\usepackage{multirow}
\usepackage{enumitem}
\usepackage{amsmath}
\usepackage[normalem]{ulem}

\usepackage{pgfplotstable}
\usepackage{tikz,pgfplots}
\usepgfplotslibrary{groupplots}
\usetikzlibrary{patterns}
\usepgflibrary{patterns}

\usepackage{listings}

\lstdefinelanguage{Relay}
{
  morekeywords={
    fn,
    let,
    Tensor,
  },
  sensitive=true, 
  morecomment=[l]{//}, 
  morecomment=[s]{/*}{*/}, 
  morestring=[b]" 
}

\definecolor{codegreen}{rgb}{0,0.6,0}
\definecolor{codegray}{rgb}{0.5,0.5,0.5}
\definecolor{codepurple}{rgb}{0.58,0,0.82}
\definecolor{backcolour}{rgb}{0.95,0.95,0.92}
\definecolor{eclipseBlue}{RGB}{42,0.0,255}
\definecolor{eclipsePurple}{RGB}{127,0,85}

\lstdefinestyle{mystyle}{
    language={Relay},
    commentstyle=\color{codegreen},
    keywordstyle=\color{magenta},
    numberstyle=\tiny\color{black},
    stringstyle=\color{codepurple},
    basicstyle=\ttfamily\footnotesize,
    emph={shape_of, alloc_storage, alloc_tensor, invoke_mut, invoke_shape_func},
    emphstyle=\color{eclipseBlue},
    breakatwhitespace=false,
    breaklines=true,
    captionpos=b,
    keepspaces=true,
    numbers=left,
    numbersep=4pt,
    showspaces=false,
    showstringspaces=false,
    showtabs=false,
    tabsize=2
}
\usepackage{hyperref}

\usepackage{color}
\usepackage{graphicx}
\usepackage{alltt}
\usepackage{lineno}
\usepackage{proof}
\usepackage{verbatim}
\usepackage{xcolor}
\usepackage{tikz}
\usepackage{xspace}
\usepackage{fancyvrb}

\newcommand{\system}[0]{\emph{Nimble}\xspace}

\newcommand{\any}{\texttt{Any}}
\definecolor{sampldark}{RGB}{72, 71, 135}
\definecolor{uwpurp}{RGB}{51, 0, 111}
\definecolor{uwpurplight}{RGB}{94, 0, 204}
\newcommand{\kwd}[1]{\textcolor{uwpurplight}{\texttt{#1}}}

\usepackage{graphics}

\usepackage{todonotes}

\newcommand{\is}[2][]{&::=&#2\ifthenelse{\equal{#1}{}}{}{&\text{(#1)}}}

\newcommand{\alt}[2][]{\\&&\vert&#2\ifthenelse{\equal{#1}{}}{}{&\text{(#1)}}}

\newcommand{\altSpace}[3]{\\[#1]&&\vert&#3\ifthenelse{\equal{#2}{}}{}{&\text{(#2)}}}

\newcommand{\rLet}[4]{\kwd{let}\ #1\ifthenelse{\equal{#2}{}}{}{\ \kwd{:}\, #2}\ \kwd{=}\ #3 \kwd{;}\ifthenelse{\equal{#4}{}}{}{\ #4}}

\makeatletter

\newcommand{\intercCheckNext}[2]{\@ifnextchar\bgroup{\intercGobbleNext{#1}{#2}}{}}
\newcommand{\intercGobbleNext}[3]{#1 #3\@ifnextchar\bgroup{\intercGobbleNext{#1}{#2}}{#2}}
\makeatother

\newcommand{\squishlist}{
   \begin{list}{$\bullet$}
     { \setlength{\itemsep}{0pt}      \setlength{\parsep}{0pt}
       \setlength{\topsep}{3pt}       \setlength{\partopsep}{0pt}
       \setlength{\leftmargin}{1em} \setlength{\labelwidth}{1em}
       \setlength{\labelsep}{0.5em} } }

\newcommand{\squishnone}{
   \begin{list}{}
     { \setlength{\itemsep}{0pt}      \setlength{\parsep}{0pt}
       \setlength{\topsep}{3pt}       \setlength{\partopsep}{0pt}
       \setlength{\leftmargin}{0em} \setlength{\labelwidth}{1em}
       \setlength{\labelsep}{0.5em} } }

\newcommand{\squishlistcirc}{
   \begin{list}{$\circ$}
     { \setlength{\itemsep}{0pt}      \setlength{\parsep}{0pt}
       \setlength{\topsep}{3pt}       \setlength{\partopsep}{0pt}
       \setlength{\leftmargin}{1.5em} \setlength{\labelwidth}{1em}
       \setlength{\labelsep}{0.5em} } }

\newcommand{\squishlisttwo}{
   \begin{list}{$\bullet$}
     { \setlength{\itemsep}{0pt}    \setlength{\parsep}{0pt}
       \setlength{\topsep}{0pt}     \setlength{\partopsep}{0pt}
       \setlength{\leftmargin}{2em} \setlength{\labelwidth}{1.5em}
       \setlength{\labelsep}{0.5em} } }

\newcommand{\squishend}{
     \end{list} \vspace{5pt} }

\newcommand{\rnum}[1]{\uppercase\expandafter{\romannumeral #1\relax}}

\usepackage{ifthen}
\newcommand{\showcomments}{yes}
\newcommand\zhicomment[1]{
    \ifthenelse{\equal{\showcomments}{yes}}{{\color{red} [zhi: #1]}}{\ignorespaces}
}
\newcommand\haichen[1]{
    \ifthenelse{\equal{\showcomments}{yes}}{{\color{magenta} [haichen: #1]}}{\ignorespaces}
}
\newcommand\yida[1]{
    \ifthenelse{\equal{\showcomments}{yes}}{{\color{blue} [yida: #1]}}{\ignorespaces}
}
\newcommand\mli[1]{
    \ifthenelse{\equal{\showcomments}{yes}}{{\color{red} [mu: #1]}}{\ignorespaces}
}
\newcommand\update[1]{
\ifthenelse{\equal{\showcomments}{yes}}{{\color{red}#1}}{\ignorespaces}
}

\newcommand\hide[1]{}

\usepackage{stfloats}

\usepackage{lipsum}

\begin{document}


\twocolumn[
\mlsystitle{Nimble: Efficiently Compiling Dynamic Neural Networks for Model Inference}
\mlsyssetsymbol{equal}{*}

\begin{mlsysauthorlist}
\mlsysauthor{Haichen Shen}{equal,aws}
\mlsysauthor{Jared Roesch}{equal,octo}
\mlsysauthor{Zhi Chen}{aws}
\mlsysauthor{Wei Chen}{aws}
\mlsysauthor{Yong Wu}{aws} \\
\mlsysauthor{Mu Li}{aws}
\mlsysauthor{Vin Sharma}{aws}
\mlsysauthor{Zachary Tatlock}{uw}
\mlsysauthor{Yida Wang}{aws}
\end{mlsysauthorlist}

\mlsysaffiliation{aws}{Amazon Web Services}
\mlsysaffiliation{octo}{OctoML}
\mlsysaffiliation{uw}{University of Washington}

\mlsyscorrespondingauthor{Haichen Shen}{shaichen@amazon.com}

\mlsyskeywords{Deep learning compiler, Dynamic neural network}

\vskip 0.1in

\begin{abstract}
  Modern deep neural networks increasingly make use of features such as control flow, dynamic data structures, and dynamic tensor shapes.
  Existing deep learning systems focus on optimizing and executing static neural networks which assume a pre-determined model architecture and
  input data shapes---assumptions that are violated by dynamic neural networks.
  Therefore, executing dynamic models with deep learning systems is currently both inflexible and sub-optimal, if not impossible.
  Optimizing dynamic neural networks is more challenging than static neural networks; optimizations must consider all possible execution paths and tensor shapes.
  This paper proposes \system, a high-performance and flexible system to optimize, compile, and execute dynamic neural networks on multiple platforms.
  \system handles model dynamism by introducing a dynamic type system, a set of dynamism-oriented optimizations, and a light-weight virtual machine runtime.
  Our evaluation demonstrates that \system outperforms existing solutions for dynamic neural networks
  by up to 20$\times$ on hardware platforms including Intel CPUs, ARM CPUs, and Nvidia GPUs.
\end{abstract}
]

\printAffiliationsAndNotice{\mlsysEqualContribution}


\section{Introduction} 
\label{sec:intro}

As deep learning-based applications have become ubiquitous, so have systems for optimizing, executing, and deploying such applications. A number of systems research projects focus on enhancing the performance of a subset of pre-trained models produced by deep learning (DL) researchers on various platforms~\cite{Dahl2011taslp, han2016isca, NIPS2016johnson, liu2019optimizing, wang2019unified}. 
Specifically, these models represented as static data flow graphs where the sizes of each input and output (i.e., tensors or $n$-dimensional arrays) are known a priori, ensuring the execution path remains unchanged on every invocation. 
We refer to models with this static nature as \emph{static models}.
Continued advances in neural networks, especially those in natural language processing, have introduced new dynamism in models, such as control flow \cite{lstm, language_model}, dynamic data structures \cite{tree_lstm, graph_lstm}, and dynamic shapes \cite{devlin2018bert}. We refer to models exhibiting these behaviors as {\em dynamic models}. 

As dynamic models mature and continue to move from research to production, it calls for an efficient and cross-platform inference system. 
This poses new challenges for deep learning practitioners, as dynamic models introduce input-dependent graph topology, breaking existing system assumptions and invalidating optimizations designed for purely static data flow graphs.
However, no existing solutions fulfill these requirements.

Many existing approaches to dynamic model optimization apply or extend existing deep learning frameworks~\cite{xu2018cavs, gao2018low, yu2018dynamic, jeong2018improving, jeong2019janus, neubig2017dynet, tensorflowfold}. 
However, deep learning frameworks optimized for training can be limiting in model inference settings due to their rich feature set. In order to realize these features frameworks are often monolithic, large, and non-portable.
Moreover, approaches which inherit from frameworks rely on third-party kernel libraries such as OpenBLAS~\cite{xianyi2014openblas}, cuDNN~\cite{cudnn}, and oneDNN~\cite{intel2020onednn} to achieve competitive performance. These libraries expose a fixed set of operators for the corresponding hardware, compromising the portability of dynamic models which require a large number of operators with varying data types and shapes. Designing a new interface independent of existing frameworks provides a clean programming model but often at the cost of performance, due to dynamic interpretation of the model~\cite{neubig2017dynet}.

An alternative approach that has generated significant interest in both academia and industry is the end-to-end optimization of neural networks using deep learning compilers, such as XLA \cite{xla}, Glow \cite{glow}, TVM \cite{tvm_osdi18}, and MLIR \cite{lattner2021mlir}. 
Deep learning compilers differ from traditional deep learning frameworks by separating execution into a compilation, and a runtime phase. The compilation phase enables whole-model optimization at the graph level, and workload specific kernel code-generation for multiple hardware platforms, while the runtime executes the compiled module.

However, deep learning compilers have been primarily restricted to static models due to lack of support for dynamism.
Specifically, in order to compile and execute the dynamic models, a system requires an intermediate representation (IR) which can statically represent dynamic constructs, a code generator 
which can generate kernels for dynamically varying data shapes, and a runtime to handle the dynamic execution and kernel dispatch accordingly.
Further, dynamic-specific optimizations, such as dynamic memory planning, the process of statically optimizing dynamic allocations, are necessary to achieve desirable performance.
None of these features exist in the current deep learning compilers.

To this end, we present \system, a high-performance and portable system for compiling, optimizing, and executing dynamic neural networks on multiple platforms.
To the best of our knowledge, this is the first attempt to systematically handle dynamic models from a compiler perspective.
First, we introduce type system extensions to handle data with unknown dimension, which is common in dynamic models, by performing type checking and inference for shapes with {\em Any}. 
Second, we devise several optimizations specific to dynamic models, including dynamic shape-aware code generation, memory planning, and device placement. 
Third, we propose a virtual machine (VM)-based runtime, which decouples the platform-independent controlling logic and platform-dependent kernel implementation, to be portable, light-weight, and most importantly, able to execute dynamic models.
Evaluation on LSTM~\cite{lstm}, Tree-LSTM~\cite{tree_lstm} and BERT~\cite{devlin2018bert} shows that \system~lowers the latency by 1.05$\times$ to 19.9$\times$ compared to the best solution whichever on mainstream hardware platforms both in the cloud (Intel CPUs and Nvidia GPUs) and at the edge (ARM CPUs).
 
In summary, this paper makes the following three core contributions:
\vspace{-1em}
\squishlist
    \item Proposes and builds an end-to-end system for efficient dynamic model inference across multiple hardware platforms, including an empirical study to benchmark the results;
    \item Devises several compilation and optimization techniques,
    including a dynamic type system, a memory planning pass, a heterogeneous device placement mechanism to place computation and data, and a symbolic kernel code generation and shape-based dispatch algorithm;
    \item Designs and implements tensor based abstract machine with a platform-independent instruction set to efficiently and flexibly execute dynamic models across platforms.
\squishend
\vspace{-1em}

The rest of the paper is organized as follows. \autoref{sec:overview} reviews the limitation of existing deep learning compilers and gives the overview of \system. 
\autoref{sec:compliation} presents the design and implementation of the compilation flow of \system, followed by VM-based runtime in \autoref{sec:runtime}. \autoref{sec:eval} provides the evaluation results using various models on different hardware platforms. 
\autoref{sec:relwk} covers related work, and \autoref{sec:conclusion} concludes the paper.

\section{Challenges and Our Approach}
\label{sec:overview}
\subsection{Limitation of Deep Learning Compilers}
\label{sec:background:dlc}
As aforementioned, existing solutions to dynamic models either rely on or extend deep learning frameworks. These solutions bring significant challenges in portability and cross-platform support due to the gigantic codebase and the vendor library dependency. Deep learning compilers provide an alternative approach as being portable across platforms with minimal memory footprint by virtue of being light-weight and dependency-free.

However, current deep learning compilers are not able to process dynamic models due to missing the following dynamism-specific features.

\squishlist
\item \textbf{An IR for representing dynamism.} Performing data type and shape inference on static models is straightforward as they are known during declaration and remain unchanged during runtime. However, the shape of an input tensor may vary wildly across different input samples in a dynamic model. The emergence of control flow constructs further complicates this problem as different execution paths can emit substantially different data. A fully static IR, hence, is inadequate to cope with the dynamic characteristics of these models.

\item \textbf{A set of dynamic-oriented optimizations.} Existing deep learning compilers, e.g. TVM \cite{tvm_osdi18} and Glow \cite{glow}, expect static input for each optimization. The memory spaces of each tensor are pre-allocated and their live cycles are determined using a dedicated optimization pass. They also ensure the homogeneous execution of the entire model because all kernels are executed on the same device. 
However, these optimizations may completely break when dynamism appears, in which different execution paths possibly require different amounts of memory with undetermined sizes before runtime. Therefore, certain IR nodes may be introduced to help runtime type inference and memory allocation. The operations in these nodes are intrinsically more CPU friendly, which would lead to the serious performance problem if not placed correctly.

\item \textbf{A symbolic kernel code generator.}
Code generation (codegen) is responsible for generating high-performance executable kernels for operators. Recent research \cite{tvm_osdi18, chen2018learning, zheng2020flextensor, adams2019learning, zheng2020ansor} has achieved impressive results in kernel performance with static shapes on multiple backends.
Nonetheless, challenges in codegen with symbolic shapes remain unexplored.
After applying the same set of loop optimization, kernels generated with symbolic shapes could still perform bad if the loop boundary is not handled properly.
Meanwhile, kernel tuning under symbolic shape settings becomes more challenging as the search space grows exponentially.


\item \textbf{A light-weight and cross-platform runtime.} For efficiency purpose, the runtime of static models could be simply designed as a sequential executor that traverses the input data flow graph in the topological order to invoke operators one by one. However, the execution path of dynamic models may only be known at runtime and the kernels for certain operators must be dispatched according to the data shape determined at runtime, making a simple graph-traversing runtime insufficient.
\squishend


\begin{figure}[t]
  \centering
  \includegraphics[width=\linewidth]{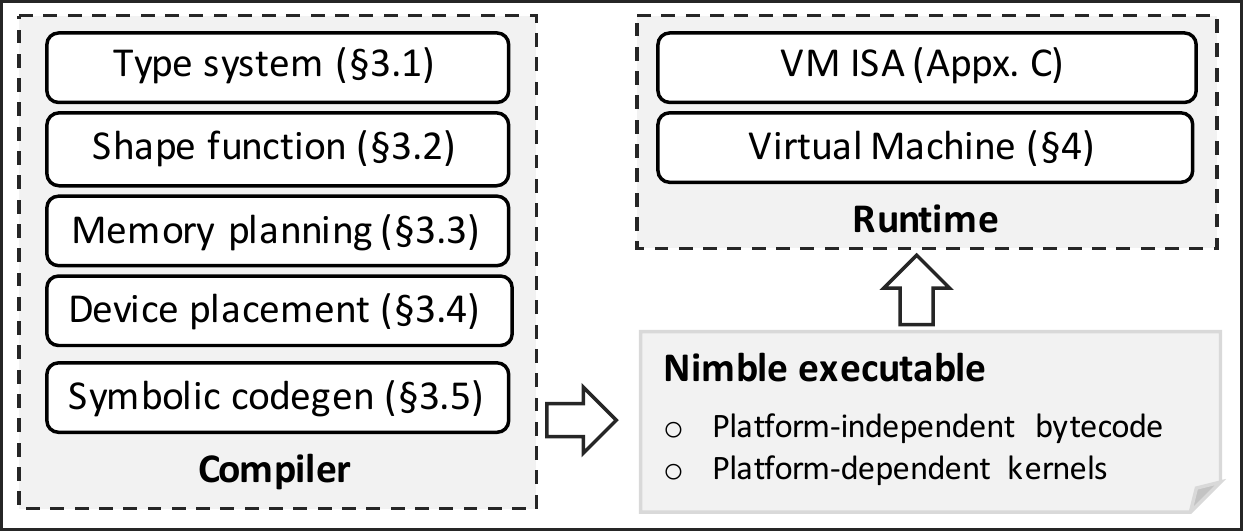}
  \captionof{figure}{\system overview. \system consists of a compiler that handles model with dynamism and a runtime that executes dynamic models in multiple platforms. 
  }
  \label{fig:vm}
  \vspace{-1.5em}
\end{figure}

\vspace{-1em}
\subsection{Our Approach}
\label{sec:background:approach}
To address these challenges, this paper presents \system, a high-performance and flexible system for compiling and optimizing dynamic models for multiple platforms. In general, the design goals of \system are:
\vspace{-8pt}
\begin{enumerate}[leftmargin=*]
    \itemsep 0em
    \item {\bf Supporting dynamic models.} \system targets models with all types of dynamism, including control flow, dynamic data structures and varied data shapes.
    \item {\bf Being portable and light-weight.} The module that \system produces should be executable across a number of platforms on the cloud (high-end CPUs and GPUs) and at the edge (low-power CPUs and GPUs). The runtime should be light enough to run on devices with minimal compute power and memory capacity.
    \item {\bf Enabling high performance.} \system should be performant in the context of dynamism across platforms.
\end{enumerate}
\vspace{-8pt}

Figure~\ref{fig:vm} shows the system architecture of \system that we propose to achieve the aforementioned design goals. 
It is a system consisting of two major components, namely a compiler and a runtime. 
\system takes a model in the format of mainstream deep learning frameworks, converts it into a unified intermediate representation (\textit{IR}), then optimizes and compiles the IR into an executable that contains both platform-agnostic bytecode and platform-dependent kernel code, and finally loads the executable to execute in the VM-based runtime. 
The bytecode is executed by \system's runtime interpreter, which is shareable across various platforms. 
This design effectively enables us to only maintain one version of the execution logic, but focus more on the performance critical operator kernels. 
The kernels are optimized for a specific hardware platform to achieve high performance. 

To effectively support dynamic models without performance degradation for static models, we introduce various analysis and optimization techniques in \system's compiler. 
First, a set of IR extensions are devised to represent dynamic shapes (\textit{Any} shape) and dynamic allocations for static optimization of dynamic program behaviors (\autoref{sec:compliation:typing}). 
Second, shape functions are attached to operators to compute the output shapes dynamically and perform type checking at runtime (\autoref{sec:compilation:shape-func}). 
Third, a memory planning optimization is employed to reduce the amount of memory consumed (\autoref{sec:compliation:memory}). 
Fourth, a heterogeneous device placement mechanism is designed to place IR nodes on ``the-best'' device to reduce expensive cross-device data transferring and synchronization (\autoref{sec:compliation:hetero}). 
Finally, the compiler features a code generator that is capable of specializing the codegen of certain likely shapes (\autoref{sec:compliation:codegen}). Once the executable with dynamic behavior is compiled, the VM-based runtime can load and interpret it with intelligent dynamic kernel dispatching (\autoref{sec:runtime}). We detail the design and implementation of each of these features in the subsequent sections.
\section{Compiler Support for Dynamism}
\label{sec:compliation}

A key challenge preventing existing deep learning compilers from handling dynamism is the lack of a uniform and dynamic representation. For example, optimizations and runtime of existing IR, e.g., TVM, always assume the presence of static shape information. These assumptions introduce quite a few challenges for optimization to dynamism. 

In order to handle dynamism, we design a set of IR extensions which expose the essential semantics required to optimize dynamic programs. The approach is implemented in \system on top of the Apache TVM (version 0.6) deep learning compiler infrastructure \cite{tvm_osdi18} to leverage its frontend converters from various DL frameworks to its IR. TVM's frontends alleviate the need to frontend specific details enabling our work to focus on contributions such as IR extensions and optimizations. To use \system, one only needs to feed it with a pre-trained model, perform compilation and then inference. Furthermore, the lessons 
here are applicable to other compiler efforts.

This section describes how we transform standard TVM programs into our dynamic dialect to easily apply static optimizations to dynamic programs, much as we do in the traditional compiler optimization. Particularly, we detail three key components required to compile dynamic models.

\vspace{-1em}
\squishlist
    \item An extended type system which enables static tracking of dynamic shapes.
    \item A series of optimization passes that make dynamic output shapes, allocation, and device placement explicit.
    \item A set of codegen techniques for producing code of kernels with dynamic input and output shapes.
\squishend

\vspace{-1em}
\subsection{Typing}
\label{sec:compliation:typing}
Deep learning compilers use type systems to represent, check and infer the data types and shapes of tensors. In some frameworks and compilers this is separated into two steps, shape inference and data type inference. TVM performs both simultaneously and refers to them as type inference~\cite{roesch2019relay}, a terminology we will use throughout the section.

A {\em Tensor Type} is designated by an $n$-dimensional shape (defined as a tuple of integers describing the tensor's dimensions) and a data type (e.g. \texttt{float32} or \texttt{int64}). 
Current deep learning IRs only support codegen when all dimensions of a tensor's shape are known at compile-time, i.e., static shapes are mandatory for type inference and checking.

In the context of dynamic models, many data shapes can only be determined at runtime. Therefore, the previous assumption of static data shapes does not hold. 
In order to support dynamic data shapes, \system introduces a special dimension called \any~to represent statically unknown dimensions. 
For example, we can represent a tensor type as \texttt{Tensor[(1, 10, Any), float32]},
where the size of the third dimension in this tensor is unknown while the other
two dimensions have concrete values.
This concept has been introduced in other frameworks. 
Janus \cite{jeong2019janus} uses similar denotation to represent a dynamic dimension but only for type unification, while \system extends type inference to handle \any~as described next. 

\noindent
{\bf Operator Type Relation} A type relation describes the relationship between types of operator inputs and outputs. The type system of TVM relies on these type relations \footnote{\tiny Relay \cite{roesch2019relay} includes more details of the type relations with static cases.} to infer and bidirectionally propagate type and shape relationships between inputs and outputs of operators across the whole network.

The type relation must be generalized to properly handle dynamic shapes. For example, a program which grows a tensor on each loop iteration (a case existing in the decoder of many NLP models) is both impossible to type and compile without proper type system support. With the introduction of \any, we are able to improve the existing type relations to support dynamic models. 

There are two dynamic type relation cases in particular.
First, operators with dynamic output shapes depending on the input data, such as {\tt arange}\footnote{\tiny{\tt arange} generates a range of values in a (start, stop, step) interval the arrays output size is a function of input arguments.} and {\tt unique}\footnote{\tiny{\tt unique} selects the unique elements of a tensor.}, will use \any~to describe those shapes. 
Second, when input shapes contain \any~dimension, the type relation needs to propagate \any~correctly to the output types and relax typing constraints that hold in the static cases when necessary.
For example, the rules
for broadcast\footnote{\tiny\url{https://docs.scipy.org/doc/numpy/user/basics.broadcasting}} type relation between two dimensions with \any~are defined as follows:
\vspace{-0.5em}
\begin{align*}
  \textrm{broadcast\_rel}(\any, 1) &\rightarrow \any \\
  \textrm{broadcast\_rel}(\any, d) &\rightarrow d ~~~~~~(d > 1) \\
  \textrm{broadcast\_rel}(\any, \any) &\rightarrow \any
  \vspace{-1em}
\end{align*}
Note that due to the presence of dynamic shapes, these type relation rules can no longer rule out all type errors at compile-time. 
For example, for the second rule shown above, when \any~is neither 1 nor $d$ at runtime, it then violates the broadcast type constraints.
To address this, we take the gradual typing~\cite{gradualtyping} approach and leave certain type checking at runtime after \any~is instantiated by a 
concrete value (see \autoref{sec:compilation:shape-func} for more details). 
One could eliminate these errors using a more advanced type system, but at increased complexity.

\noindent
{\bf Type Inference} 
One caveat of the \any~dimension is that unknown dimensions will propagate during type inference, 
    reducing chances for shape specialization. 
To limit the loss of precision introduced by using \any~dimensions,  we introduce {\em sub-shaping} to the type system.
Much like sub-typing used in popular programming languages \cite{LiskovTPLS1994,AmadioAmadioTPLS1993}, 
    our type system extension enables values with concrete dimensions to be valid sub-types of 
    tensor types with dynamic dimensions.
For example a \verb|Tensor[(128, 128), f32]| is a valid sub-type of \verb|Tensor[(any, 128), f32]|.
Furthermore we use program analysis to detect and remove unnecessary dynamism, by communicating extra
    shape information which can be used in downstream compilation.
We do this in two critical spots:
    first we refine any false dynamic dimensions with static dimensions using a secondary shape analysis;
    and second, in code generation we use a single variable dimension for equivalent dynamic dimensions.

\subsection{Shape Function}
\label{sec:compilation:shape-func}
The introduction of \any~dimension invalidates the pre-allocation mechanism adopted in the existing deep learning compiler.
Instead, we now have to track the amount of memory required to be allocated in parallel to computing. 
Furthermore, static type checking cannot eliminate all type errors at compile-time due to dynamic tensor shapes.
Consequently, we define a {\em shape function} to compute the output shape for storage allocation and verify the type relation in accord with the semantics of every operator. 
The shape function is similar in structure to the type relations described in \autoref{sec:compliation:typing} but are present at runtime instead of compile-time. 
It enables compiling and embedding the computation of output shapes into the program.

Based on the characteristics of operators, we divide the shape functions in three different modes: data independent, data dependent, and upper bound. 
{\em Data independent} shape functions are used for operators in which the output shape only depends on the shapes of inputs such as normal 2-D convolution. 
{\em Data dependent} shape functions require the concrete input values to compute the output shapes. For example, the output shape of \texttt{arange} depends on the value of start, stop, and step. 
In addition, there are certain operators such as Non Maximum Suppression (\texttt{nms}) where the complexity of computing the output shapes is on par with the complexity of executing the operator itself. 
To avoid the redundant computation, we use an {\em upper bound} shape function to quickly estimate an upper bound shape for the output. 
We require such operators to return the output shape along with output value, so as to use the real shape to slice the output tensors into precise output shape and layout. 

It is worth noting that in the presence of dynamic shape functions, operator fusion needs to be specially taken care of. 
Operator fusion, which combines {\em basic operators} into a {\em composite operator}, is a critical technique for performance optimization as it reduces unnecessary memory copies and improves the cache locality. 
The compiler can easily connect the shape functions of basic operators to form the shape function for a composite operator when all shape functions are data independent.
However, a basic operator with a data dependent or upper bound shape function cannot be fused to other operators, i.e., taking the outputs of other operators as its inputs to fuse together, as the shape function requires to access to the intermediate result within a composite operator.
As a result, we explicitly define the fusion policy to prevent this from happening.


\subsection{Memory Planning}
\label{sec:compliation:memory}

Many deep learning compilers use a form of static memory planning that
    coalesces memory into contiguous chunks and minimizes allocations. 
For devices such as GPUs these optimizations are essential for 
    reducing memory fragmentation and ensuring allocation does not hamper kernel performance. 
Existing deep learning compiler IRs hide memory allocation behind a 
    functional interface, where each operator implicitly allocates its output storage.
Before execution, the system then performs static memory planning on the data 
    flow graph enabling efficient pre-allocation of the output buffers. 
Due to this ``out-of-band'' nature of memory allocation, 
    it is challenging to customize, modify, or compose memory optimizations with other passes. 
For example, if one needs to adjust memory allocation for heterogeneous execution, 
    modifications to the runtime are required. 
TVM's graph runtime is one such example of static memory planning.
Due to the coarse-grained memory semantics of deep learning models, 
    it is essential that memory optimizations occur at a suitably high-level of abstraction, 
    unlike traditional compilers.
Existing work provides no clear path to performing static optimization on dynamic memory
    allocation. 

In order to perform what we refer to as ``dynamic memory planning'' we have extended 
    TVM to transform its IR with implicit memory allocations to one with explicit 
    buffer allocation and manipulation. 
The key to this transformation is an inter-procedural change of calling convention, 
    with each operator now taking its outputs explicitly. 
The transformation makes it possible to track and optimize dynamic memory allocations in the IR.
In order to perform this optimization we have introduced four new IR constructs, 
    (a) \verb|invoke_mut(op, inputs, outputs)| which takes outputs as mutable in-out arguments, 
    (b) \texttt{alloc\_storage(size, alignment, device)} which allocates a region of memory of a particular size, 
    (c) \texttt{alloc\_tensor(storage, offset, shape, dtype, attrs)} which allocates a tensor at a particular storage offset with a shape and data type, and 
    (d) \verb|kill(tensor)| which frees a tensor before its reference count becomes zero due to exiting the frame.

We illustrate how this works with an example of transforming a single statically shaped operation such as broadcasting addition. 
Note that in the below code examples \texttt{Tensor<d1, ..., dn>} is shorthand for a tensor of shape \texttt{(d1, ..., dn)} containing floating point values.

\begin{lstlisting}
fn main() -> Tensor<10> {
  let t1, t2 : Tensor<10> = ...;
  add(t1, t2)
}
\end{lstlisting}
\vspace{-1em}

Here we only must allocate a single buffer, that is, the return buffer for the addition operation.

\begin{lstlisting}
fn main() -> Tensor<10> {
  let t1 = ...; let t2 = ...;
  let buf1 = alloc_storage(40,64,cpu);
  let out1 = alloc_tensor(buf1,0,(10),f32);
  invoke_mut(add, (t1, t2), (out1));
  out1
}
\end{lstlisting}
\vspace{-1em}

The above transformation replaces the operator invocation \verb|add| to code which 
first allocates an output tensor from backing storage at offset zero, and call to 
\verb|invoke_mut|. 
For the sake of space, we present a more complex example in the \autoref{appx:mem-plan-example}, which illustrates how to handle memory allocation when operators have dynamic shaped inputs.
The key insight is to internalize a notion of memory allocation into the IR, enabling static optimization of 
both static and dynamic allocations in the presence of control and dynamic shapes. 
Now that all allocations are explicit in the IR, we can provide analogous optimizations in the static
case on dynamic programs, for example we have implemented a storage coalescing pass to group storage
into a larger region which allows the multiplexing of multiple tensor allocations to a single piece of storage.
Futher optimization like liveness analysis and graph coloring algorithm can be applied to the program to reuse  the storages.

\subsection{Heterogeneous Device Placement}
\label{sec:compliation:hetero}
As discussed in \autoref{sec:compilation:shape-func}, shape functions are executed at runtime to calculate the output shape of an operator. These functions should execute on the CPU as their outputs are used to compute the size of allocated memory. In the case of heterogeneous execution (i.e., CPU and GPU), it is essential to carefully place the execution of IR nodes to proper devices. Otherwise, considerable overhead from data transfers and device synchronization will occur if the inputs to shape functions and kernels need to be copied from or to GPU. To minimize the performance penalty, we analyze the program to place sub-expressions on the most suitable devices.

\begin{figure}
    \centering
    \includegraphics[width=\linewidth]{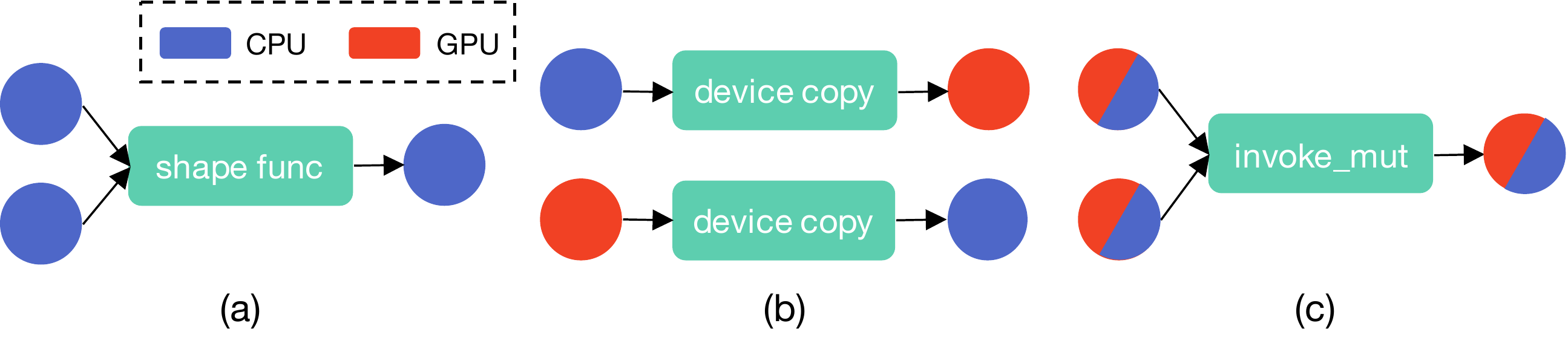}
    \caption{Some heterogeneous device placement rules. (a) The inputs and outputs of shape functions are placed on CPU. 
    (b) \texttt{device\_copy} changes the device of output accordingly. 
    (c) The device of all arguments to \texttt{invoke\_mut} must be the same.
    }
    \label{fig:hetero}
    \vspace{-2em}
\end{figure}

We introduce a unification based analysis for computing the correct device placement and allocation based on the previous scheduling of the compute kernels. The goal of our device analysis is to assign each IR node properly to minimize the number of cross-device copies. We introduce a concept of \texttt{DeviceDomain} to represent the domain of a device, including source and destination. Each expression in the IR defaults to the empty domain, meaning no constraint on its device placement. In addition, a new IR construct, \verb|device_copy|, is introduced to facilitate the heterogeneous execution of the \system runtime. It represents a data transfer between different devices and is inserted when a cross-device data copy is mandatory. 
Our analysis is formulated as a set of device placement rules which describes how device constraints flow, and then we use unification, a technique common in type inference and compilers, to compute the precise device placement. 
\autoref{fig:hetero} depicts some highlights of the rules to assign and propagate the device types: (a) both inputs and outputs of shape function are assigned to CPU, (b) the output of \verb|device_copy| is assigned to the device that is copied to, (c) all arguments to \verb|invoke_mut| should have the same device domain. The full set of rules can be found in the \autoref{appx:hetero-rules}.

Based on these rules, we use a union-find data structure to bidirectionally propagate and unify the device placement of each IR node. We introduce two operations, \texttt{union(s, t)} and \texttt{find(s)}, to achieve \texttt{DeviceDomain} unification throughout the entire program. \texttt{union(s,t)} unions the equivalence device domains of \texttt{s} and \texttt{t} into one equivalence domain when the device types match. \texttt{find(s)} returns the representative of the device domain that \texttt{s} belongs to. These two operations are applied until all IR nodes are annotated. 
If there is no constraint to a device domain, we assign the compilation target (i.e., GPU) to it in favor of better kernel performance.
The result of the heterogeneous device placement composes with memory planning and shape function insertion resulting in correctly placed allocations.

\subsection{Symbolic Codegen}
\label{sec:compliation:codegen}
Deep learning compilers~\cite{tvm_osdi18, halide} have demonstrated competitive performance compared to manually tuned kernels on multiple platforms. Recent trends apply machine learning based search to further reduce or eliminate complex manual performance tuning using either template based~\cite{chen2018learning, zheng2020flextensor} or search based~\cite{adams2019learning, zheng2020ansor} approaches.
However, existing work which focuses on tuning in the presence of static shapes falls short with symbolic or dynamic shapes. There are two inherent challenges with regard to codegen of symbolic shapes.
\vspace{-.5em}
\squishlist
    \item How to achieve the same performance of kernels generated with symbolic shapes as that with static shapes when applying the same schedule?
    \item How to extend the machine learning based approach to tune kernels with symbolic shapes?
\squishend
\vspace{-1em}

Loop parallelism and loop tiling are common optimization techniques that exploit multi-core capabilities by achieving data access patterns which are memory hierarchy aware for both CPUs and GPUs. However, the combination of these techniques lead to complex loop boundary conditions. In static cases, we can prove these conditions always hold, and thus eliminate checks that hamper further optimizations such as unrolling. 
While straightforward to handle with static shapes, it becomes a non-trivial challenge when performing symbolic codegen. If not carefully handled, the boundary condition checks will stay, leading to poor performance. 

To address this issue, we generate multiple kernels according to the residues modulo of the tiling factor and then dispatch based on the actual shape at runtime. 
For example, suppose a symbolic dimension $x$ is tiled by a factor of 8, we then duplicate the generated kernel for 8 times, and replace the symbolic var $x$ by $8k+r$ in each copy, where $k = \lfloor x / 8 \rfloor$ and $r \in [0..7]$. By applying this technique in conjunction with an enhanced symbolic expression simplification pass, we can eliminate most boundary checks to achieve performance that is nearly identical to kernels compiled with a single static shape. Lastly, we automatically generate a dispatch function that invokes the corresponding kernel based on the residue. 
In addition, the dispatch function can be extended to invoke either compiler generated kernels or {\em third party library} whichever is faster from the profiling results.
The increased kernel size is relatively small compared to the overall deep learning models.
In case where resources are extremely limited, we can either generate fewer number of kernels than the tiling factor or reduce the tiling factor to find an acceptable trade-off between code size and performance.

A known issue to machine learning based tuning is that it takes a long time (usually hours) to find the best schedule for a single kernel. When it comes to symbolic shapes, the tuning time can be exponentially longer if we naively tune for every possible shape. In this paper, we extend the template based tuning approach for symbolic shapes to make tuning time tractable. The template based tuning approach takes a human-defined code template and a search space, and searches the best configuration within the search space using machine learning algorithms.
We observe that a good configuration for one shape usually performs well on other shapes. Based on this observation, we devise the following mechanism to tune the kernel for symbolic shapes.

\vspace{-12pt}
\begin{enumerate}[leftmargin=*]
    \itemsep 0em 
    \item First replace the symbolic dimensions by a large enough value (e.g., 64) so that the search space can cover most possibilities, and run the tuning algorithm on the static shape for a sufficient number of iterations.
    \item Pick top $k$ configurations, apply them to a selection of other shapes, and evaluate their performance.
    \item Pick the configuration that performs best on average among shapes previously evaluated.
\end{enumerate}
\vspace{-12pt}

Empirically, we found that $k=100$ covers most of the best configurations for other shapes. Current popular dynamic models usually only require kernels with one symbolic variable. As a result, we choose the values of power of two up to 256 in the cross evaluation of other shapes. If there is more than one symbolic variable, a more sophisticated selection approach might be required to limit the evaluation time of step 2. We leave this to the future work. Further, if the workload distribution is known, we can adjust the weighting of known shapes in step 3.

\section{Virtual Machine}
\label{sec:runtime}
The conventional runtime of existing deep learning compilers which naively executes a model operator by operator in topological order does not work for executing the compiled modules of dynamic models. A more intelligent and powerful execute engine is required to handle the control flow execution logic, and dispatch different kernels accordingly. In order to achieve these goals and be portable to different platforms, we design and implement a virtual machine (VM)-based runtime.

In \system, we compile a dynamic model into a {\em VM executable} that contains platform-independent bytecode and platform-dependent kernel code, which can be later loaded and executed. 
The bytecode consists of a series of instructions that predicate the order of kernel invocation and control flow execution logic.
This design compliments conventional runtime's capability for executing highly optimized kernels but not directly handling orchestration between kernels. 


The design of the VM instruction set is motivated by the simple observation that kernel execution dominates neural network execution time. 
It is quite different from traditional language virtual machines, which contain many instructions that perform little work, leading to a profile where the cost of each instruction executed matters. 
Our ISA is composed of CISC-style instructions in which each instruction corresponds to a primitive IR expression on tensors, such as allocation and kernel invocation, which in turn may correspond to executing multiple ``low-level'' operations. For example, \texttt{LoadConst idx, \$reg} is capable of multiple addressing modes as it first reads the index \texttt{idx} and then loads the data from a constant pool to the destination register \texttt{\$reg}.
A complete list of instruction set can be found in the \autoref{appx:vm}.
We naturally select a register-based virtual machine design~\cite{davis2003case} for a compact bytecode, which is easy for users to read and modify. We provide the abstraction of an infinite set of virtual registers as it significantly simplifies optimizations and allocation (similar to SSA) and minimizes conceptual barriers to rapid prototyping and modification.
 
Instructions are represented using a traditional tagged union containing the op-code and the data payload. This representation enables both efficient serialization and instruction decoding and dispatch. \system~uses variable-length instruction format due to the inclusion of variable sized operands such as data shapes in the instructions.

After we have generated an executable from the compiling phase,
we can create an interpreter to load it. When execution begins, the interpreter runs a dispatch loop which checks the op-code and executes the appropriate logic, then repeats. As our instructions are coarse-grained (i.e., they can be viewed as super-instructions), the number of branches generated by the dispatch-loop is lower than traditional programming language VMs, adding negligible overhead compared to ahead of time compilation.

\noindent {\bf Discussion} 
An alternative solution to the runtime is ahead of time compilation 
to eliminate dispatch overhead.
But due to the granularity of the operations, dispatch time makes up a very small portion in the execution. More importantly, our VM provides flexibility traditionally attributed to virtual machines and a clear compiler/runtime split. 
We see the potential of VM to be integrated as a runtime module into a larger system.
For example, VM can provide resource isolation where multiple inference instances share the same hardware in the cloud. Furthermore, a Quality of Service (QoS)-aware system, e.g., \cite{kang2018hotmobile, Yachir2009rsj}, could leverage VM to suspend the current model execution for a higher priority or time-critical model. Last, thanks to the simplicity of the VM design, one can verify the implementation of VM for security and privacy purposes.

\section{Evaluation}
\label{sec:eval}
This section evaluates the performance of \system on dynamic models against existing state-of-the-art solutions, as well as its optimization implication. Specifically, the section seeks to answer the following questions:
\squishlist
    \item What is the overall performance of \system for dynamic models compared against state-of-the-art alternatives on various hardware platforms?
    \item How much overhead does \system VM introduce for handling dynamism at runtime?
    \item How effective are the proposed optimization techniques, such as memory planning and symbolic codegen?
\squishend

\vspace{-1em}
\subsection{Experiment setup}
\label{sec:eval:setup}

All experiments were conducted on Amazon EC2 instances. We evaluated \system on three hardware platforms: Intel Skylake CPUs (c5.9xlarge, 18 physical cores, hereinafter called {\em Intel CPU}), Nvidia Tesla T4 GPUs (g4dn.4xlarge, 1 card, 2,560 CUDA cores, hereinafter called {\em Nvidia GPU}), and ARM Cortex A72 (a1.4xlarge, 16 physical cores, hereinafter called {\em ARM CPU}). Although all tests are done on the cloud, our results of ARM CPU are portable to the edge devices, e.g. Raspberry Pi, due to the same architecture. 

To study the efficiency of \system in handling dynamic models, we compared it with mainstream deep learning frameworks, including TensorFlow (v1.15), MXNet (v1.6), PyTorch (v1.5) \footnotemark, DyNet (v2.1), as well as dynamic-specific systems TensorFlow Fold based on TensorFlow v1.0.
\footnotetext{\tiny We use PyTorch v1.4 on ARM CPU because PyTorch v1.5 fails to build on ARM instance.}
We were unable to compare \system with Cavs~\cite{xu2018cavs}, JANUS~\cite{jeong2019janus}, or Jeong et al.\cite{jeong2018improving} as none of them is open-source. No public deep learning compiler has claimed support for dynamic models.

Three popular models that represent different classes of dynamism were chosen in this experiment, viz. LSTM~\cite{lstm} (dynamic control flow), Tree-LSTM~\cite{tree_lstm} (dynamic data structure), and BERT~\cite{devlin2018bert} (dynamic data shape).
The input size / hidden size used in the LSTM and Tree-LSTM model are 300/512 and 300/150, respectively.
We used BERT base implementation.
For LSTM and BERT, we used Microsoft Research's Paraphrase Corpus (MRPC)~\cite{dolan2005microsoft} with variable input lengths as our input dataset. For Tree-LSTM, we used the Stanford Sentiment Treebank (SST)~\cite{socher2013recursive} with various tree structures as the input dataset.

\vspace{-.5em}
\subsection{Overall performance}
\label{sec:eval:overall}
We compared the overall performance of \system against baselines for each dynamic model. \system successfully accomplished inference for all models on all platforms.
However, not all baseline systems could perform inference for these models.
For instance, Tree-LSTM only runs on PyTorch, DyNet, and TensorFlow Fold as other frameworks cannot handle dynamic data structures. TensorFlow Fold was not designed to process BERT hence no result was obtainable. We cannot find a BERT implementation on DyNet.
Finally, the model inference of Tree-LSTM on Nvidia GPU was omitted as it's hard to saturate GPU compute capability due to excessive control flows and small kernel sizes, making GPU less favorable deployment targets.

The baseline systems all make use of third-party kernel libraries to achieve high-performance by leveraging the heavily hand-optimized operators, which is handicapped when an operator is not supported on a specific target. 
However, \system has the ability to select either the self-compiled kernels or the ones provided by third-party library based on which one maximizes performance.
It uses dynamic dispatch logic to invoke the selected kernels using platform-independent bytecode at runtime. 

First, the latency result comparison is shown in \autoref{tab:lstm}. \system consistently outperforms the baseline on both 1- and 2-layer cases, with $2.2\times$, $1.3\times$, and $3.2\times$ faster than the best alternative on Intel CPU, Nvidia GPU, and ARM CPU, respectively.
We implemented the LSTM model using a for-loop control flow in all systems for fair comparison.
PyTorch has an alternative implementation for LSTM that unrolls the LSTM cells along the sequence length and dynamically batches the matrix multiplication for the inputs. Note that this optimization is orthogonal to the control flow handling and can be implemented in \system.
DyNet is significantly slow on CPU because it only utilizes a single core.
We observe that latency on Nvidia GPU is higher than Intel CPU with \system. This is because the size of LSTM model is relative small so that it cannot fully utilize the massive parallelism in the GPU.
The significant performance improvement of \system comes from two aspects: (a) utilizing the deep learning compiler for better kernel fusion and implementation, (b) encoding the control flow into platform-independent instructions with minimal overhead.

\begin{table}[t]
\centering
\small
\begin{tabular}{p{0.9cm}|ccc|ccc}
\toprule
Unit: & \multicolumn{3}{c|}{1 layer} & \multicolumn{3}{c}{2 layers} \\
$\mu$s/token & Intel & NV & ARM & Intel & NV & ARM \\ \midrule
\system & \bf{47.8} & \bf{54.6} & \bf{182.2} & \bf{97.2} & \bf{107.4} & \bf{686.4} \\
PT & 103.6 &  80.6 & 2735.9 &  224.0 & 158.1 &  5862.2 \\
DY & 936.4 &  68.8 & 5701.6 & 2350.8 & 140.7 & 12811.3 \\
MX & 212.9 & 135.7 & 3695.9 &  401.7 & 223.8 &  7768.0 \\
TF & 301.4 & 304.7 &  978.3 &  687.3 & 406.9 &  2192.8 \\
\bottomrule
\end{tabular}
\caption{LSTM model inference latency of \system, PyTorch (PT), DyNet (DY), MXNet (MX), and TensorFlow (TF) on Intel CPU, Nvidia (NV) GPU, and ARM CPU.}
\label{tab:lstm}
\vspace{-1.2em}
\end{table}

Next, we inspected the performance of model inference on Tree-LSTM as exhibited in \autoref{tab:treelstm}. The table shows that \system runs substantially faster than the baselines. On PyTorch, the performance speedups are $17.4\times$ on Intel CPU and $19.8\times$ on ARM CPU as PyTorch uses Python to handle the tree data structure.
DyNet performs much better than PyTorch as it handles the control flow execution carefully in C++. However, \system still outperforms it as DyNet does not optimize computation kernels for CPUs.
TensorFlow Fold is $5.2\times$ slower than \system on Intel CPU because it has to re-compile upon every input.\footnotetext{\tiny TensorFlow Fold was not built successfully on ARM CPU.}

\begin{table}[t!]
\centering
\small
\begin{tabular}{l|ll}
\toprule
Unit: $\mu$s/token        & Intel     & ARM \\ \midrule
\system  & \bf{40.3}  & \bf{86.3}  \\
PyTorch & 701.6 & 1717.1  \\
DyNet & 98.2 & 312.9 \\
TF Fold \footnotemark & 209.9 & --  \\
\bottomrule
\end{tabular}
\caption{Tree-LSTM model inference latency.} 
\label{tab:treelstm}
\vspace{-1.5em}
\end{table}

Last, \autoref{tab:bert} summarizes the performance comparison of BERT. The results indicate that \system outstrips the baselines for all frameworks on all platforms in the experiment. The reduction in latency compared to the best alternative is $1.5\times$, $1.3\times$, and $1.05\times$ on Intel CPU, Nvidia GPU, and ARM CPU, respectively.
The reasons are two-fold: (a) similar to frameworks, \system is also able to use the well-tuned third-party libraries on Intel CPU (MKL) and Nvidia GPU (cuDNN); (b) \system can leverage more thorough operator fusion brought by the deep learning compiler. 

In sum, the evaluation results demonstrate that \system produces more portable performance for all dynamic models on different platforms. Instead, the performance of frameworks is more platform dependent and varies from model to model.

\begin{table}[t]
\centering
\small
\begin{tabular}{l|lll}
\toprule
Unit: $\mu$s/token    & Intel  &   Nvidia       &  ARM     \\ \midrule
\system     & \bf{307.0} & \bf{95.2} & \bf{2862.6} \\
PyTorch & 479.5 & 220.4 & 11851.2 \\
MXNet      & 455.8 & 152.9 & 8628.0   \\
TensorFlow & 768.7 & 125.2 & 2995.4 \\
\bottomrule
\end{tabular}
\caption{BERT model inference latency.} 
\label{tab:bert}
\vspace{-1.5em}
\end{table}



\vspace{-.2em}
\subsection{Microbenchmark}

\begin{table}[t]
    \centering
    \small
    \begin{tabular}{c|cc|cc}
        \toprule
        \multirow{2}{*}{Device} & TVM & \system  & kernel & others  \\
        & lat. (ms) & lat. (ms) & lat. (ms) &  (ms) \\ 
        \midrule
        Intel  & 19.38 & 24.32 & 21.06 & 3.26 \\
        ARM & 223.50 & 237.41 & 228.59 & 8.82 \\
        Nvidia & 5.58 & 5.86 & 5.60 & 0.26 \\
        \bottomrule
    \end{tabular}
    \caption{BERT model latency (sequence length 128) using TVM and \system on different hardware.
    {\it kernel latency} shows the time of kernel execution in \system, and {\it others} shows the extra latency introduced by other instructions.
    }
    \label{tab:overhead}
    \vspace{-1.5em}
\end{table}

\hide{
\begin{figure}[t]
    \centering
    \includegraphics[height=3.5cm]{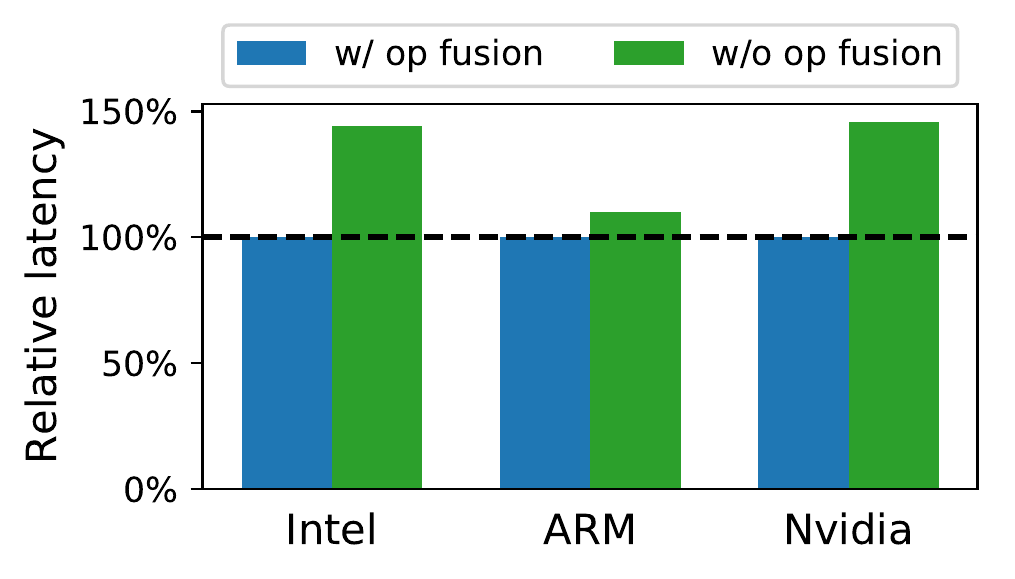}
    \caption{Relative latency comparison between \system with and without operator fusion for BERT model with sequence length 128.}
    \label{fig:fusion}
\vspace{-1em}
\end{figure}
}

\begin{figure}[t]
    \centering
    \includegraphics[height=4.3cm]{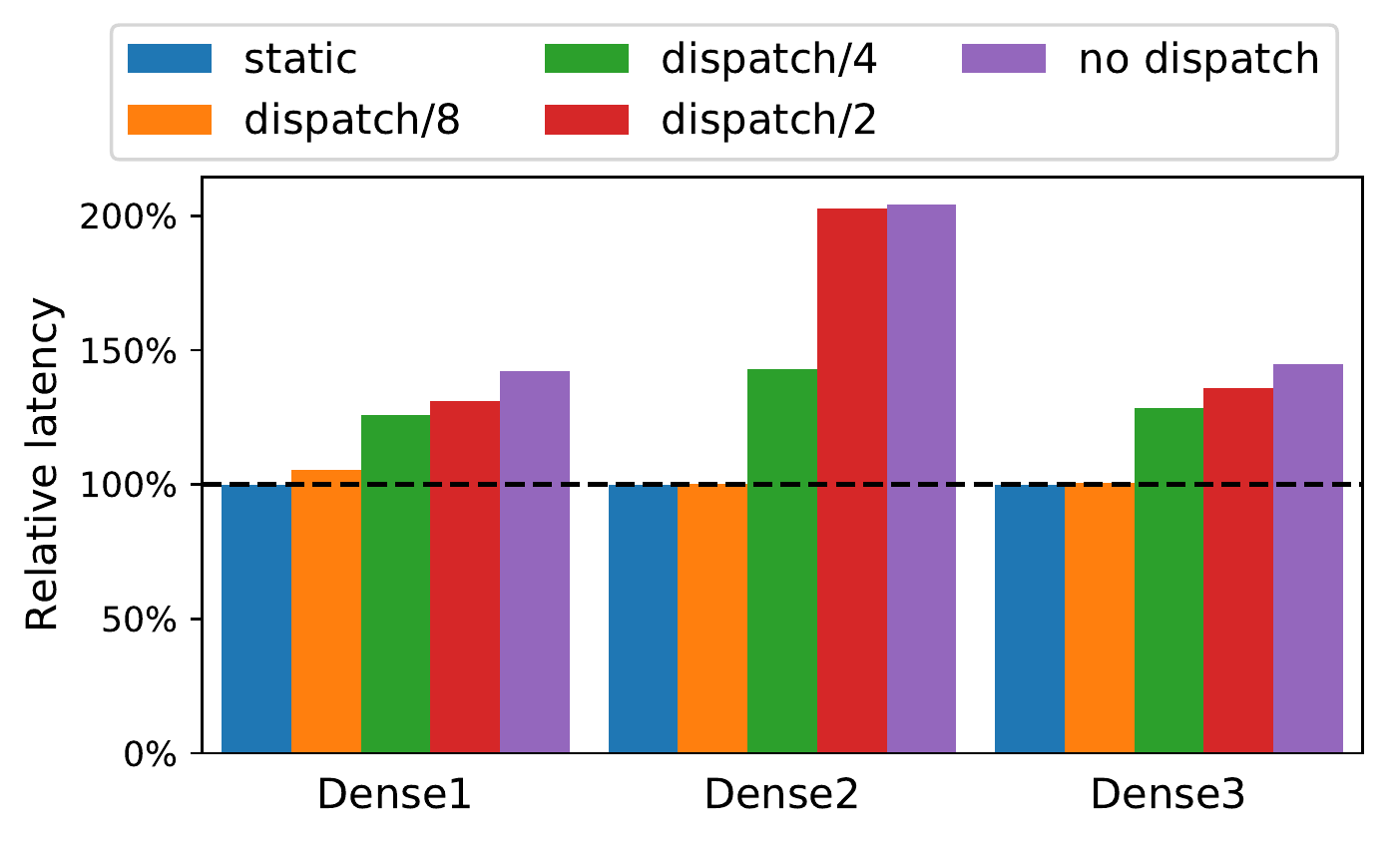}
    \caption{Relative latency of 3 dense operators using symbolic codegen and static codegen on ARM CPU. The latency of static-shaped kernels is used as the baseline. ``dispatch/$k$'' means that we generate $k$ symbolic kernels to be dispatched at runtime. ``no dispatch'' means that only one symbolic kernel is generated and therefore no dispatching is needed. 
    \vspace{-1.5em}
    }
    \label{fig:sym-codegen}
\end{figure}

This subsection analyzes the performance gain of \system by using BERT as the microbenchmark. Three studies will be conducted to examine (a) the overhead introduced by the VM, 
(b) the advantage of the proposed memory planning pass, and (c) the performance discrepancy between symbolic and static codegen. Other models share similar observations.

\noindent {\bf Overhead in handling dynamism} In order to understand the overhead that \system spends to take care of dynamism, we compared it to TVM where static sequence length and TVM static runtime is used to execute BERT.
\autoref{tab:overhead} details the performance difference between \system and TVM.  TVM is 5\% to 25\% faster than \system on static shapes, though the absolute latency difference is small. The overhead comes from two aspects: (a) kernels generated with symbolic shapes cause extra overhead in the index computation; (b) other instructions in the VM are required to handle the dynamic execution, such as shape functions, dynamic memory allocation, instruction dispatch, etc.
On Nvidia GPU, most of the bytecode latency is overlapped with the GPU execution thanks to the heterogeneous device placement (\autoref{sec:compliation:hetero}), and therefore the overhead of other instructions is negligible.

\hide{
\noindent {\bf Operator fusion}
One of the most important benefits from deep learning compiler is its flexibility to fuse operators for better cache locality. \autoref{fig:fusion} depicts the comparison between \system with and without operator fusion. With operator fusion, the latency can be reduced by 44\%, 10\%, and 46\% on Intel CPU, ARM CPU, and Nvidia GPU, respectively. This is because operator fusion provides better opportunities for (a) achieving better cache locality, and (b) reducing the buffer allocation for temporary intermediates as well as the number of shape functions to be executed.\yida{Mention that this is not available if we don't take care of the shape functions properly. Op fusion is not our contribution, but enabling it in the context of dynamism is.}}

\noindent {\bf Memory planning} 
\autoref{sec:compliation:memory} proposed memory planning to coalesce memory allocation together and reuse the already allocated memory chunks. This pass reduces the number of buffer allocation by 47\%, and the memory allocation latency is reduced by 75\% 
on Intel CPU. We also compared the memory usage of \system with memory planning to TVM which statically analyze and pre-allocate memory on popular computer vision models such as ResNet~\cite{he2016deep}, MobileNet~\cite{howard2017mobilenets}, VGG~\cite{simonyan2014very} and SqueezeNet~\cite{iandola2016squeezenet}. It turned out that \system uses up to 8\% more memory footprint.

\noindent {\bf Symbolic codegen} We selected 3 dense layers from BERT model and compared the performance between symbolic codegen and static codegen on ARM CPU.
For symbolic codegen, we use \verb|Any| as the sequence length during the compilation and evaluate the kernel with the sequence length 128 at runtime. For static codegen, we directly set the sequence length to 128 at compilation time.
\autoref{fig:sym-codegen} illustrates the relative latency of kernels generated with symbolic shapes to the baseline -- kernel compiled with static shapes.
The auto-tuning algorithm tiles the symbolic axis by 8 in all three kernels. We varied the number of generated kernels to be dispatched during the symbolic codegen from 8 (full dispatch) to 1 (no dispatch) as described in \autoref{sec:compliation:codegen}. We observe that symbolic codegen with full dispatch can achieve nearly identical performance to that for static codegen, while reducing the number of kernels hurts the performance.
Similar trends are seen in dense operators with different shapes, other operators, and on other platforms.

\section{Related Work}
\label{sec:relwk}

This section contrasts \system to existing solutions 
    for executing dynamic neural networks. 
    
{\bf Deep learning frameworks} 
Some frameworks support dynamic control flow via the addition of primitives in their graph representations, such as \emph{switch} and \emph{merge} in TensorFlow~\cite{yu2018dynamic} and {\em foreach}, {\em cond}, 
    and {\em while\_loop} in MXNet~\cite{mxnet-control}. 
Indirect encodings of control flow require specialized data-flow runtimes which handle operations like switch, 
    or hybrid runtimes which separate execution of the control and data planes. Both are heavily intrusive to the framework codebase.
In addition, there have been many framework extensions to support different kinds of dynamism.
TensorFlow Fold~\cite{tensorflowfold} conducts an analysis of the user's provided computation graph to identify dynamic operations that can be batched together. 
Once such operations are found, a batched TensorFlow graph is generated which
    provides shape specialized sub-graphs.
Although this may provide speedup, it introduces large overhead as each path must be 
    executed as a separated sub-computation graph, as well as limits further optimization. 
Jeong et al.~\cite{jeong2018improving} and JANUS~\cite{jeong2019janus} also 
    extend TensorFlow to improve the performance for dynamic models. 
These extensions are framework specific and are not directly related 
    to compilation techniques we employ. 
The use of speculative execution is complimentary to our techniques. 
    
Dynet~\cite{neubig2017dynet} and PyTorch~\cite{pytorch} use host language features (i.e., Python's control flow) to 
    dynamically unroll control flow to produce a static trace of a dynamic model. 
Tracing based approaches provide a flexible and friendly programming model at the cost of ahead of time optimization.
Additionally, it requires the creation of a data flow graph for each trace, introducing overhead
    for control-flow constructs and limiting whole-program optimization, a challenge also faced
    in traditional tracing JIT compilation. 
JAX~\cite{jax2018github} also supports restricted forms of dynamic networks 
    but its optimizations are fundamentally restricted by XLA, its underlying compiler.
For example, dynamic value dependent control-flow is not supported in JAX's JIT mode.
In contrast \system makes dynamic behaviors less costly to use without compromising performance for the static subset.

In addition, frameworks rely on third-party libraries~\cite{cudnn,intel2020onednn} to implement operators with 
    different data shapes, namely, they achieve good performance for models with 
    dynamic shapes on a specific hardware platform only if the corresponding high-performance third-party library supports an operation.
Therefore, frameworks, as well as the runtime systems derived from them for dynamic models~\cite{xu2018cavs, gao2018low}, generally perform poorly on devices in the second tier of support such as ARM CPU, and on operator and shape combinations not found in popular benchmarks.
In contrast \system works across all platforms and can generate performant code for new shapes and new devices on demand (Section~\ref{sec:compliation:codegen}).

{\bf Deep learning compilers} 
Existing deep learning compilers, including XLA~\cite{xla}, TVM~\cite{tvm_osdi18}, and Glow~\cite{glow}, 
    can compile deep learning models to run on multiple hardware platforms with accelerated performance. 
However, little work has been done on optimizing compilation for dynamic neural networks.
MLIR~\cite{lattner2021mlir} is a promising direction and its IR supports dynamic shapes, but no dynamic optimizations or end to end performance have been reported yet.
\system's compilation and VM design is largely inspired by production compilers and VMs, such as LLVM~\cite{llvm}, GCC~\cite{gcc}, and JVM~\cite{jvm} for general solutions to handle dynamic behaviors, such as control flow and variable-length input arrays.


\vspace{-2pt}
\section{Conclusion}
\label{sec:conclusion}
This paper proposed \system, an end-to-end compiler and runtime solution to dynamic neural networks. \system is the first deep learning compiler that supports neural networks with dynamism, via a lightweight and portable VM-based runtime for executing compiled models on multiple platforms. 
Experimental results showed that \system efficiently executed popular dynamic models on multiple platforms with better performance and broader coverage compared to the state-of-the-art. Future work includes enabling dynamic model inference on emerging AI accelerators and high-performance training of dynamic models.


\bibliographystyle{mlsys2021}
\bibliography{nimble}

\clearpage
\appendix
\section*{Appendices}

\section{Memory Planning Example}
\label{appx:mem-plan-example}

In the case of operators with dynamic shaped inputs, we need to insert shape functions before the kernel invocation to compute the output shapes and allocate memory accordingly as detailed in~\autoref{sec:compilation:shape-func}.
Our uniform treatment of shape functions as standard tensor expressions enables them to be fused and optimized like normal, but one challenge is that we must now manifest memory allocations in a fixed point until we allocate the outputs for both the compute and necessary shape functions. 
We illustrate this below how the explicit memory allocation transformation works with a single dynamic concatenation.

\begin{lstlisting}
fn (x: Tensor<?, 2>, y: Tensor<1, 2>)
    ->Tensor<?, 2> {
  concat((%x, %y)) 
}
\end{lstlisting}

This is the same transformation as the simple example shown in \autoref{sec:compliation:memory} with the addition of carefully inserting 
invocations to the shape function to compute sizes of output buffers for the dynamically sized kernel.

\begin{lstlisting}
fn (x: Tensor<?, 2>, y: Tensor<1, 2>)
    ->Tensor<?, 2> {
  let in_sh0 = shape_of(x);
  let in_sh1 = shape_of(y);
  let buf0 = alloc_storage(16, 64, ...);
  let out_sh0 = alloc_tensor(buf0, ...);
  invoke_shape_func(concat, 
      (in_sh0, in_sh1), (out_sh0,), ...);
  let buf1 = alloc_storage(...);
  let out0 = alloc_tensor(
      buf1, out_sh0, ...);
  invoke_mut(concat, (x, y), (out0));
  out_0
}
\end{lstlisting}

After the transformation you may notice we have introduced a call to \verb|invoke_shape_func| which invokes a shape function for a kernel (line 7). The shape function requires input shapes as arguments which further require us to invoke \verb|shape_of| for both \verb|%x| and \verb|%y|~(line 3-4). \verb|shape_of| will be directly mapped to a VM instruction to retrieve the shape of a tensor at runtime. The transformation also inserts an additional storage and tensor allocation for the output of the shape function (line 5-6).

\section{Heterogeneous Device Placement Rules}
\label{appx:hetero-rules}

The full set of heterogeneous device placement rules are listed below:

\begin{itemize}
    \item \verb|shape_of|. Defaults to the CPU domain because we can access the shape of a tensor regardless of which device it is placed on.
    \item Shape functions. These IRs take the output of one or multiple \verb|shape_of| and then derive the shape of an operation according to predefined type inference rules. The output of a shape function is used to compute the amount of memory that this operator requires at runtime, which only needs a few cheap scalar arithmetic computation. Therefore, the inputs and outputs would be better on a CPU domain as well.
    \item \verb|device_copy|. The input and output of this IR are on different domains as it copies data from one domain to another. The device domains of the input and output are propagated in the opposite directions to other IR nodes that are reachable to/from the device copy node.
    \item Memory operations. The device domain of storage from \verb|alloc_storage| is designated in the expression, and later is propagated to the device domain of the tensors allocated from this storage via \verb|alloc_tensor|.
    \item \verb|invoke_mut|. All arguments used in the \verb|invoke_mut| must have the same device domain.
    \item Other common IR nodes. The device domain of other common IR nodes, e.g. variables, constants, operators, etc., can be directly propagated from the above nodes.
\end{itemize}

\section{VM ISA}
\label{appx:vm}

\setcounter{table}{0}
\renewcommand{\thetable}{C.\arabic{table}}

\begin{table*}[!tp]
\centering
\begin{tabular}{ll}
\toprule
Instruction    & Description                                                                   \\ \midrule
Move           & Moves data from one register to another.                                      \\
Ret            & Returns the object in the result register to the caller's register.           \\
If             & Jumps to the true or false offset depending on the condition.                 \\
Goto           & Unconditionally jumps to an offset.                                           \\
LoadConst      & Loads a constant at an index from the constant pool.                          \\
LoadConsti     & Loads a constant immediate.                                                   \\
AllocStorage   & Allocates a storage block on a specified device.                              \\
AllocTensor    & Allocates a tensor object with a static shape from a storage.                                  \\
AllocTensorReg & Allocates a tensor object given the shape in a register.                                       \\
AllocADT       & Allocates a data type using the entries from a register.                      \\
AllocClosure   & Allocates a closure with a lowered virtual machine function.                  \\
FreeStorage    & Free allocated memory back to memory manager. \\
FreeTensor     & Release the memory occupied by a tensor back to the storage object. \\
Invoke         & Invokes a function.                                               \\
InvokeClosure  & Invokes a closure.                                                     \\
InvokePacked   & Invokes an optimized operator kernel.             \\
GetField       & Gets the value at a certain index from a VM object.                           \\
GetTag         & Gets the tag of an Algebraic Data Types (ADT) constructor.                    \\
DeviceCopy     & Copies a chunk of data from one device to another.                            \\
ShapeOf        & Retrieves the shape of a tensor.                                              \\
ReshapeTensor  & Assigns a new shape to a tensor without altering its data.                    \\
Fatal          & Raises fatal in the VM.                                                       \\
\bottomrule
\end{tabular}
\caption{The opcode and the description of \system's instruction set.}
\label{tab:isa}
\end{table*}

\autoref{tab:isa} details the opcode and the functionality of each instruction. Recall that \system is designed to support neural networks with dynamic features, such as control flow and dynamic data structures etc., in a portable, high-performance, and light-weight manner. A set of instructions are proposed to fulfill this task. These instructions provide not only high-level information about the dynamic model behavior but also an architectural level interface for better orchestration and virtualization of the execution of control logic and optimized operator kernels. The current instruction set only contains 22 instructions for dynamic model inference. It largely reduces the dispatching overhead and simplifies bytecode serialization and deserialization. We categorize these instructions as follows:

\begin{itemize}
    \item Register-to-Register Operations. Register-to-Register operations, e.g. \texttt{Move}, transfer data between different offset of the register file. Objects are reference counted, make use of copy-on-write and passed by reference ensuring register operations are cheap even if the size of underlying container is large.
    
    \item Memory Operations. Memory operations can allocate space for tensors, load constant tensors, and so on. Due to the design of our constant pool, weights (which are constant during inference) can remain in-memory with no specialized support. They can be referenced by the \texttt{LoadConst} instruction. \texttt{FreeStorage} and \texttt{FreeTensor} free the memory before the reference count becomes zero.
    
    \item Call Operations. Call operations are the most frequently executed instructions. The ISA
    has specialized call instructions for invoking a global function, a kernel primitive, and a closure. \texttt{InvokePacked} is the most performance-critical one. It is in charge of invoking the operator kernels that are optimized either by the underlying deep learning compiler or a third-party library. Kernel primitives are ahead-of-time compiled and can leverage both compiler-generated kernels and the third-party libraries using the dispatch function described in \autoref{sec:compliation:codegen}.
    On the other hand, both \texttt{Invoke} and \texttt{InvokeClosure} call into a global VM function where a closure object carries the captured registers.
    
    \item Control Flow Operations. Unconditional jump instructions, e.g. \texttt{Goto} and \texttt{Ret}, are used by both static and dynamic models to jump to a specific program point. Only dynamic models need conditional control operations, e.g. \texttt{If}, to determine the direction of branching. The interpreter updates the PC using the offset from either the true branch or false branch based on the conditional value.
    
    \item Miscellaneous Operations. To ease compiler optimizations (e.g. memory planning and device placement) and code generation, we offer the native support of several instructions in the VM, namely \texttt{ShapeOf}, \texttt{DeviceCopy}, and \texttt{ReshapeTensor}. These three instructions are used to directly manipulate runtime data, such as extracting the shape of a tensor, moving data between different devices, and transforming the shape of a tensor. With the help of them, we could preserve more coarse-grained IR at the frontend making optimization simpler.
\end{itemize}





\end{document}